\documentclass[11pt,preprint]{emulateapj}
\usepackage{amsmath,euscript,amsfonts}
\usepackage[normalem]{ulem}

\def\cfa{1}
\def\su{2}
\def\uv{3}
\def\hro{4}
\def\yu{5}

\shorttitle{Radio and X-ray Supernova 2013df}
\shortauthors{Kamble et al.}

\begin{document}

\title{Radio and X-rays from SN\,2013df Enlighten Progenitors of type IIb Supernovae}
\author{
Atish Kamble\altaffilmark{\cfa},
Raffaella Margutti\altaffilmark{\cfa}, 
Alicia ~M. Soderberg\altaffilmark{\cfa},
Sayan Chakraborti\altaffilmark{\cfa},
Claes Fransson\altaffilmark{\su},
Roger Chevalier\altaffilmark{\uv},
Diana Powell\altaffilmark{\cfa},
Dan Milisavljevic\altaffilmark{\cfa},
Jerod Parrent\altaffilmark{\cfa},
Michael Bietenholz\altaffilmark{\hro,}\altaffilmark{\yu}
}

\altaffiltext{\cfa}{Harvard-Smithsonian Center for Astrophysics,
        60 Garden St., Cambridge, MA 02138}

\altaffiltext{\uv}{Department of Astronomy, University of Virginia, P.O. Box 400325, 
        Charlottesville, VA 22904-4325, USA} 

\altaffiltext{\su}{Department of Astronomy, The Oskar Klein Centre, 
        Stockholm University, AlbaNova University Centre, SE-106 91 Stockholm, Sweden}

\altaffiltext{\hro}{Hartebeesthoek Radio Observatory, 
	PO Box 443, Krugersdorp, 1740, South Africa}	

\altaffiltext{\yu}{Department of Physics and Astronomy, 
	York University, Toronto, M3J 1P3, Ontario, Canada}
	
%---------------------------------------------------------------------------------------------------------------%
\begin{abstract}
We present radio and X-ray observations of the nearby Type IIb Supernova 2013df in NGC\,4414 from 10 to 250 days after the explosion. The radio emission showed a peculiar soft-to-hard spectral evolution. We present a model in which inverse Compton cooling of synchrotron emitting electrons can account for the observed spectral and light curve evolution. A significant mass-loss rate, $\dot{M} \approx 8 \times 10^{-5}\,M_{\odot}$/yr for a wind velocity of 10 km/s, is estimated from the detailed modeling of radio and X-ray emission, which are primarily due to synchrotron and bremsstrahlung, respectively. We show that SN 2013df is similar to SN\,1993J in various ways. The shock wave speed of SN\,2013df was found to be average among the radio supernovae; $v_{sh}/c \sim 0.07$. We did not find any significant deviation from smooth decline in the light curve of SN 2013df. One of the main results of our self-consistent multiband modeling is the significant deviation from energy equipartition between magnetic fields and relativistic electrons behind the shock. We estimate $\epsilon_{e} = 200 \epsilon_B$. 
In general for Type IIb SNe, we find that the presence of bright optical cooling envelope emission is linked with free-free radio absorption and bright thermal X-ray emission. This finding suggests that more extended progenitors, similar to that of SN 2013df, suffer from substantial mass loss in the years before the supernova. 
\end{abstract}
%---------------------------------------------------------------------------------------------------------------%

\section{Introduction}

With the recent advent of high cadence ($\lesssim 1$ day) optical
transient searches, our observational understanding of the nature of
stellar explosions has bloomed.  Through these efforts, exotic breeds
of supernovae have been revealed with peculiar light-curve and
spectral properties, often prominent in the first hours to days
following the explosion (e.g., \citealt{Kasliwal2010,Drout2013}).
The rare and intriguing class of Type IIb supernova (SN IIb) --
showing evidence for both hydrogen and helium in early spectroscopic
observations \citep{Ensman1987,Filippenko1988,Filippenko1997} -- 
has become one focus point of these new
optical transient surveys. The luminosity and evolution of the early
emission places strict constraints on the physical parameters (e.g.,
mass, radius, age, stellar wind) of the pre-explosion progenitor
system \citep{Baron1993,Filippenko1993,Swartz1993} 
as it maps to the cooling envelope phase of the
ejecta immediately following shock breakout from the
stellar surface \citep{Katz2010,Nakar2010}.

Here we present radio and X-ray observations of the
recent type IIb SN 2013df in NGC4414 to investigate the nature of its progenitor
system and the pre-explosion mass loss history.  Based on detailed
modeling of the entire dataset, we clearly show that the SN is
essentially an analogue to SN\,1993J and stems from an extended
progenitor star, in line with the results of pre-explosion
imaging\citep{Van Dyk2014} and post-explosion UV observations \citep{Ben-Ami2014}.
As in the case of SN\,1993J, our early radio and X-ray observations
point to dense material in the local environs as evidenced by inverse
Compton processes. We show that these observations imply  
diversity among the progenitors and may provide crucial 
diagnostics to distinguish between possible sub-classes of Type IIb SNe.

SN IIb were originally identified as a unique class of core-collapse
supernovae two decades ago with the discovery and long-term
multi-wavelength study of SN\,1993J in M81 \citep{Filippenko1993,van Dyk1994,
Pooley1993,Filippenko1997}.  
There is general consensus that the progenitor star of SN 1993J was a massive
K type super-giant star with $R\approx 600~R_{\odot}$ and $M\approx 13-22~M_{\odot}$
based on pre-explosion imaging \citep{Aldering1994,Cohen1995,Maund2009,Fox2014}. Thanks to its proximity, 
SN\,1993J was studied in depth and the early optical emission was observed
to {\it decline} over the first $\sim 6$ days prior to rebrightening due to
radioactive decay products.  In addition, the unusual nature
of the late-time spectral evolution proved indicative of circumstellar
interaction with dense material ejected prior to stellar collapse the
origin of which remains a topic of debate \citep{Matheson2000,van Dyk1994, Chandra2009}.  Subsequent
discoveries of SN IIb saddled them in a class intermediate between
normal H-rich Type II supernovae (SN II) and H-poor Type Ibc
supernovae (SN Ibc) based on their spectroscopic properties.   

One unifying feature and hallmark of the class of SN IIb is
circumstellar interaction as evidenced by strong non-thermal
emission in the radio and X-ray bands due to dynamical interaction of
the shockwave with the surrounding medium \citep{Chevalier1982,Chevalier2003}.
In addition to SN\,1993J, such signatures have been observed for Type IIb
SN\,2001ig \citep{Ryder2006}, SN\,2003bg \citep{Soderberg2006},
SN\,2008ax \citep{Roming2009}, SN\,2011dh \citep{Soderberg2012,Krauss2012,de Witt2015},
SN\,2011ei \citep{Milisavljevic2013}, SN\,2012au \citep{Kamble2014}. 
In roughly half of these cases, episodic flux density
modulations of the non-thermal emission are observed and point to
circumstellar medium (CSM) density fluctuations on a radial scale
$\lesssim 10^{17}$ cm \citep{Ryder2004,Wellons2012}, possibly due to
a binary companion or mass loss variability (due to stellar pulsations) 
prior to explosion.   

Combining early optical observations with radio and X-ray data for a
small sample of SN IIb, \citet{Chevalier2010} proposed that the class of SN IIb
is diverse and encompasses both ``compact'' (Type cIIb; $R\sim
R_{\odot}$) and ``extended'' (Type eIIb; $R\sim 100-1000~R_{\odot}$)
progenitor stars.  The authors suggest that compact progenitors are
distinct and can be identified based on the shockwave velocity as well
as the luminosity and duration of the optical cooling envelope phase.
In this framework, SN\,1993J was clearly identified as a SN\,eIIb
while most of the other Type IIb's were identified as being more closely aligned with
a SN cIIb classification, thereby bearing more resemblance to SN Ibc.
We note that the binary classification into either SN cIIb or SN eIIb is likely simplistic 
and it is possible that there is a continuum of properties across the SN IIb
class \citep{Horesh2013}.  

% The organization of this paper

\section{Observations}
\label{sec:obs}

	\subsection{Radio Observations using \emph{Very Large Array}}
	\label{sec:obs-rad}

	%Discovery
	SN\,2013df was optically discovered on 7.87 June 
	2013 by the Italian Supernovae Search Project, with an offset of about
	$32\farcs0$ W and $14\farcs 0$ N from the center of the host
	galaxy NGC 4414 \citep{Ciabattari2013} at a distance of
	$d\approx 16.6\pm0.4$ Mpc \citep{Freedman2001}. The spectroscopic
	classification of SN\,2013df being of type IIb was provided by 
	\citet{Ciabattari2013} on 10.8 June 2013. Based on the comparison
	of light curve evolution of SN\,2013df and SN\,1993J \citet{Van Dyk2014}
	estimate that the supernova occurred on JD 2,456,447.8 $\pm$ 0.5 
	or June 4.3. Throughout this article, we use this to be the date of explosion.
	
	We first observed SN 2013df with Jansky Very Large Array (VLA) \footnote{The Jansky 
	Very Large Array is operated by the National Radio Astronomy Observatory, a
	facility of the National Science Foundation operated under cooperative
	agreement by Associated Universities, Inc.} on 2013 Jun 14.0 UT
	at the VLA C band. We did not detect the supernova on this occasion 
	and also on our subsequent observation on 2013 June 26.1 UT, 
	about 10 and 22 days after the supernova, respectively.
	On our third attempt on 2013 July 5.9 we detected a bright radio source coincident 
	with the optical position at $\alpha\rm (J2000)=12^{\rm h} 26^{\rm m} 
	29.33^{\rm s}$ and
	$\delta\rm (J2000)=+31^{\rm o}13^{\prime}38.3^{\prime\prime}$ ($\pm\,0.1^{\prime\prime}$ in each
	coordinate) with flux density of $f_{\nu}=0.65\pm 0.05$ mJy at 8.6 GHz.
	This, and subsequent observations from 1.5 to 44 GHz, are
	summarized in Table~1.

	%Radio observations
	All observations were taken in standard continuum observing mode with
	a bandwidth of $16\times 64$ MHz. During the reduction
	we split the data in two sub-bands ($8\times 64$ MHz) of
	approximately 1 GHz each. We calibrated the flux
	density scale using observations of 3C~286,
	and for phase referencing we used the calibrator 4C 21.35
	(JVAS J1224+2122). 
	The data were reduced using standard packages within the
	Astronomical Image Processing System (AIPS).

%------------------------------------------------------------------------------------------------------------%
\begin{deluxetable}{lcccr}
\tablecaption{VLA radio flux density measurements of SN\,2013df\tablenotemark{a}}
\tablewidth{0pt}
\tablehead{ \colhead{Date}	& \colhead{MJD}	&	\colhead{Frequncy}	& \colhead{$F\pm \sigma$\tablenotemark{b}}	& \colhead{Array} \\
\colhead{(UT)} 				& \colhead{} 		& 	\colhead{(GHz)} 	& \colhead{(mJy)} 		& \colhead{Config.}}
\startdata
%2013\,Jun\,14.00&	56457.00	&	4.8	& 	0.039	$\pm$	0.021	&      	C	\\
%\ldots                  &	\ldots  	&	7.1	&    	-0.024	$\pm$	0.017	&	\ldots	\\
%2013\,Jun\,26.07&	56469.07	& 	4.8	& 	0.010	$\pm$	0.027	& 	C	\\	
%\ldots                  &	\ldots  	&	7.1	& 	0.050	$\pm$	0.023	&	\ldots	\\
2013\,Jun\,14.00&	56457.00	&	4.8	& 	$<0.063$					&      	C	\\
\ldots                  &	\ldots  	&	7.1	&    	$<0.051$					&	\ldots	\\
2013\,Jun\,26.07&	56469.07	& 	4.8	& 	$<0.081$					& 	C	\\	
\ldots                  &	\ldots  	&	7.1	& 	$<0.069$					&	\ldots	\\
2013\,Jul\,05.86&	56478.86	&	8.6	&   	0.651	$\pm$     	0.046	&	C		\\
\ldots           	&	\ldots  	&	11.0	&    	0.732	 $\pm$    	0.034	&	\ldots	\\
\ldots              	&	\ldots	&	13.5	&    	0.647	$\pm$	0.041	&	\ldots	\\
\ldots              	&	\ldots  	&	16.0	&    	0.615	$\pm$	0.033 	&	\ldots	\\
2013\,Jul\,21.09&	56494.09	&	13.5	&   	2.107    	$\pm$ 	0.030	&	C		\\
\ldots              	&	\ldots	&	16.0	&    	1.928    	$\pm$ 	0.029	&	\ldots	\\
\ldots              	&	\ldots	&	19.2	&    	1.934    	$\pm$ 	0.031	&	\ldots	\\
\ldots              	&	\ldots	&	24.5	&    	1.406     	$\pm$ 	0.043	&	\ldots	\\
\ldots              	&	\ldots	&	30.0	&    	0.939  	$\pm$ 	0.044	&	\ldots	\\
\ldots              	&	\ldots	&	43.7	&    	0.351   	$\pm$ 	0.092	&	\ldots	\\
2013\,Aug\,11.11&	56515.11	&	5.0	&   	1.552    	$\pm$ 	0.043	&	C	\\
\ldots             	&	\ldots 	&	7.1	&     	2.639   	$\pm$ 	0.041	&	\ldots	\\
\ldots             	&	\ldots	&	13.5	&    	2.356    	$\pm$ 	0.035 	&	\ldots	\\
\ldots             	&	\ldots	&	16.0	&    	2.145   	$\pm$ 	0.035	&	\ldots	\\
2014\,Jan\,27.36&	56684.36	&	1.5	&   	1.531   	$\pm$ 	0.100 	&	BnA	\\
\ldots             	&	\ldots	&	5.0	&     	2.303    	$\pm$ 	0.023	&	\ldots	\\
\ldots             	&	\ldots	&	7.1	&     	1.748    	$\pm$ 	0.020	&	\ldots	\\
\ldots             	&	\ldots	&	13.5	&    	0.909   	$\pm$ 	0.023	&	\ldots	\\
\ldots             	&	\ldots	&	16.0	&    	0.773    	$\pm$ 	0.022	&	\ldots	\\
...             		&	\ldots	&	33.5	&    	0.279   	$\pm$ 	0.038	&	\ldots	
%------------------------------------------------------------------------------------------------------------%
\enddata
\tablenotetext{a}{The quoted upper limits are 3$\sigma$}
\tablenotetext{b}{image rms}
%\tablenotetext{a}{The flux errors mentioned here correspond to the image RMS. 
%	Systematic uncertainties due to external factors such as weather, calibrators etc. 
%	have not been included in the table but have been taken into consideration during modeling.}
\label{tab:vla}
\end{deluxetable}
%------------------------------------------------------------------------------------------------------------%

\subsection{Radio Spectral Energy Distribution of SN 2013df}
The overall spectral energy distribution (SED) of SN\,2013df appears 
to be brightening with time at least until the observations of August 11, 2013
or within first 70 days after the SN. The peak spectral luminosity ($d=16.6$ Mpc) 
$L_{\nu_{p}} = 3 \times 10^{26}\, \rm erg/s/Hz$ on 11 August, 2013 is about 4 times 
brighter than it was for the first detection only a month earlier on 2013 July 5.9.
Interestingly, the spectral index $\beta$ (for $f_{\nu} \propto \nu^{\beta}$) 
is rather steep at early times ($\beta = -1.6\pm0.2$ on 21st July 2013) and 
becomes shallower as it ages ($\beta = -0.9\pm0.1$ on 27th January 2014).

This trend is unusual for the radio emission from SN. As the SN shock wave 
sweeps through the circumstellar medium (CSM) around a massive star,  
it shock accelerates the CSM electrons which generates the synchrotron emission.
For the wind density profile ($n(r) \propto r^{-2}$) found around massive stars
the total amount of matter swept up by the shock wave in a unit distance, 
or equivalently in unit time for a non-decelerating shock-wave, remains the same.
As a result, the constant spectral peak brightness routinely observed in radio 
supernovae is the hallmark of the wind density profile. The evolving spectral 
index is also suggestive of a radiation process in addition to the synchrotron
emission and is evidently predominant at early times. These properties are considered 
in detail below.

%------------------------------------------------------------------------------------------------------------%
	\subsection{X-ray Observations using \emph{Swift} and \emph{Chandra}}
	\label{sec:obs-Xray}

%------------------------------------------------------------------------------------------------------------
\subsubsection{Swift-XRT}
\label{SubSec:XRTObs}

\begin{figure*}
\vskip -0.0 true cm
\centering
\includegraphics[width=8cm]{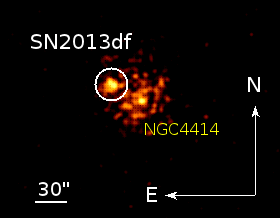} 
\includegraphics[width=8cm]{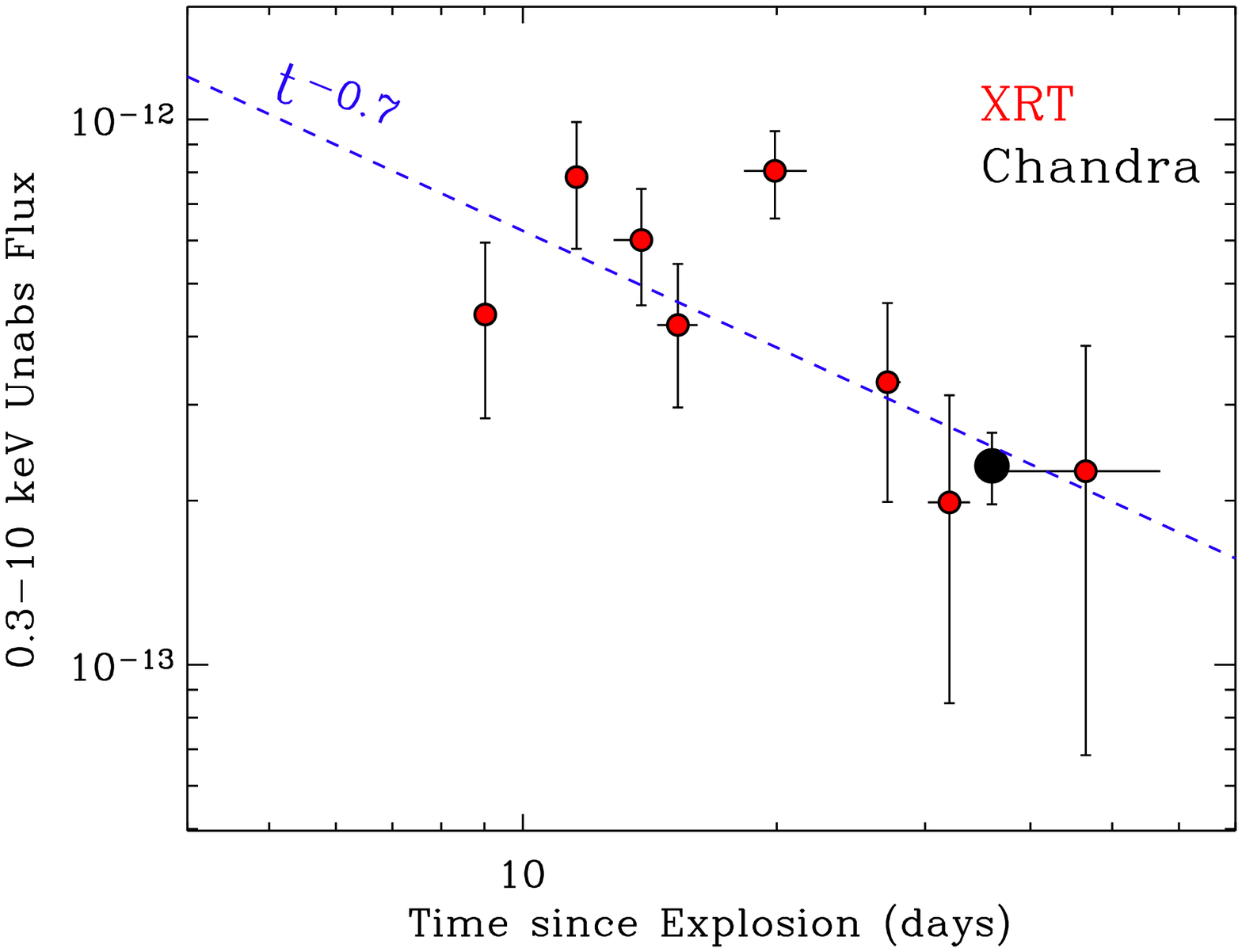} 
\caption{\emph{Left panel}: X-ray emission from SN\,2013df is clearly detected by the Swift-XRT (0.3-10 keV). The image comprises observations collected starting June 13, 2013 until August 6, 2013, total of $49.5$ ks  (i.e. $\sim8-60$ days since explosion).
\emph{Right panel}: The X-ray emission from SN\,2013df as captured by the \emph{Swift}-XRT and Chandra X-ray Observatory (CXO) shows a temporal decay consistent with a power-law $t^{-\alpha}$ with best-fitting $\alpha=0.71\pm0.16$ (1 $\sigma$ c.l.).}
\label{fig:XRTtotal}
\end{figure*}

The X-Ray Telescope (XRT, \citealt{Burrows2005}) onboard  the \emph{Swift} satellite \citep{Gehrels2004} 
started observing SN\,2013df on 2013 June 13 ($\delta t=8$  days since explosion, 
PI Roming). We analyzed the XRT data using  HEASOFT (v6.15) and corresponding calibration files.
Standard filtering and screening criteria were applied.  A fading X-ray source is clearly detected at the position of SN\,2013df that we associate with the SN shock interaction with the medium (Fig. \ref{fig:XRTtotal}).  

Figure \ref{fig:XRTtotal} shows the temporal evolution of the X-ray emission from SN\,2013df. We find a decay consistent with a power-law $\propto t^{-\alpha}$ with $\alpha=0.71 \pm0.16$ (1 $\sigma$ c.l.). The contribution of diffuse X-ray emission from the host galaxy to the  \emph{Swift}-XRT data has been estimated and subtracted by using the \emph{Chandra} X-ray observations below (section \ref{SubSec:ChandraObs}) where the point-like emission from SN\,2013df is well resolved.  

A combined spectrum made from the entire \emph{Swift}-XRT data set is well modeled by an absorbed power-law with hard photon index $\Gamma=1.58 \pm 0.06$ ($1\sigma$) and no evidence for neutral hydrogen absorption intrinsic to the host galaxy, in agreement with the results from the spectral analysis of \emph{Chandra} observations which we discuss in the next section \ref{SubSec:ChandraObs}. The Galactic neutral hydrogen column density in the direction of SN\,2013df is $N_{H_{MW}}=1.60\times10^{20}\,\rm{cm^{-2}}$ \citep{Kalberla2005}. 
%------------------------------------------------------------------------------------------------------------------
\subsubsection{Chandra}
\label{SubSec:ChandraObs} 

We initiated deep X-ray follow up of SN\,2013df with the \emph{Chandra} X-ray  Observatory
on 2013 July 10, corresponding to $\delta t=36$ days since explosion (PI Soderberg). Data have been reduced
with the CIAO software package (version 4.6) and corresponding calibration files. Standard ACIS data filtering has been applied.  In this observation which lasted for 9.9 ks we clearly detected SN\,2013df, with significance $>50\,\sigma$. As for the \emph{Swift}-XRT observations, the spectrum is well modeled by an absorbed power-law with a hard photon index $\Gamma=1.26\pm0.25$ ($1\sigma$). We again find no evidence for intrinsic neutral hydrogen absorption, with a $3\sigma$ upper limit of $N_{H_{int}}<2\times10^{21}\,\rm{cm^{-2}}$. The complete X-ray light-curve of SN\,2013df comprising \emph{Swift}-XRT and \emph{Chandra} observations is presented in Figure \ref{fig:XRTtotal}. 

\begin{figure}[h]
\vskip -0.0 true cm
\centering
\includegraphics[scale=0.4]{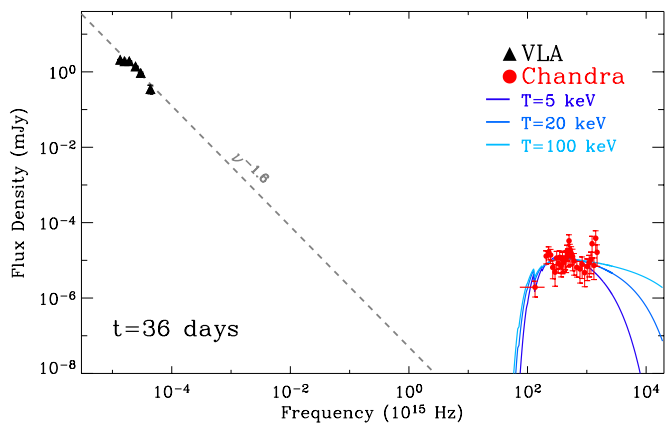}
\caption{Radio to X-ray SED of SN\,2013df at the time of the \emph{Chandra} observations. The X-ray flux is clearly in excess to the synchrotron model that best fits the radio observations (grey dashed line) and can be explained by thermal bremsstrahlung emission with temperature $T>5$ keV. Observations at $E>10$ keV such as might be done by \emph{NuSTAR} is necessary to constrain the temperature of the emitting plasma.}
\label{fig:XraySED} 
\end{figure}

The hard X-ray spectrum can be fit by a thermal bremsstrahlung model with $kT > 5$ keV (Figure \ref{fig:XraySED}). Due to relatively small number of counts and lack of coverage in hard X-ray band, we could establish only a lower limit on the plasma temperature. 

\section{Inverse Compton Model}
\label{sec:IC}
The main radiation process powering radio SNe is synchrotron emission. 
A shock wave is an efficient way to convert bulk kinetic energy into thermal
motion of the shock accelerated particles. The SN shock wave accelerates 
electrons to relativistic speeds which then gyrate in the shock amplified magnetic 
field and radiate synchrotron radiation. The electrons are usually assumed to have a power-law 
energy distribution, $n_{e}(\gamma_{e}) \propto \gamma^{-p}_{e}$, with index $p$
and Lorentz factor $\gamma_{e}$ of individual electrons as a proxy of electron 
energy. This form of the electron distribution is evident from the synchrotron 
radio spectrum.

Two radiative cooling processes can affect this particle distribution: 
synchrotron cooling and inverse Compton scattering (IC). The synchrotron, as well as 
IC, power of an individual electron is strongly dependent on its Lorentz factor, 
$P_{\nu} \propto \gamma^2_{e}$. As a result, high energy electrons 
tend to cool down faster. 
The cooling timescale for electrons emitting synchrotron radiation 
at frequency 
$\nu$ is given by \citet{Bjornsson2004} as
\begin{equation}
t_{sc} = 1.7 \times 10^{2} B^{-3/2} \nu^{-1/2}_{10} \rm days
\end{equation}
when $\nu_{10} \equiv \nu/10^{10}$ Hz \footnote{Throughout this paper 
we will follow the notation $X_{a} = X/10^{a}$, with quantities expressed in cgs units}. 
Thus, the synchrotron cooling will be dominant only when $t_{syn}$ is the smallest 
timescale in the system i.e. $t_{sc} \ll (t,t_{ic})$.

The particle distribution could also be affected when high energy electrons
cool down by scattering off optical photons due to photospheric emission 
from the SN. In the process, electrons cool down by imparting a fraction 
of their energy to the photons. The cooling time scale for this process 
depends on the reservoir of optical photons. 
For a SN having bolometric luminosity $L$ and size $r$
the average photon energy density would be given by $U_{ph} = L/4\pi r^{2} c$.
Using this the cooling time scale is given by 
\begin{equation}
t_{ic} = 1.5 \frac{B^{1/2} ~r^{2}_{15}} {L_{42} ~\nu^{1/2}_{10}}~\rm days
\end{equation}

There is another way of demonstrating the importance of IC as the dominant 
cooling process over synchrotron emission. It can be shown 
that the cooled down electrons have a steeper Lorentz factor distribution 
$n_{e}(\gamma_{e}) \propto \gamma^{-(p+1)}_{e}$ which is similar 
in the case of IC and synchrotron cooling, since both the processes have 
similar dependence on electron Lorentz factor:
$P_{\nu} \propto \gamma^2_{e} (U_{B} + U_{ph})$. If the local
magnetic field energy density ($U_{B}$) dominates over the local
photon energy density ($U_{ph}$), synchrotron cooling will be stronger
than IC cooling and vice-versa. The effective electron distribution, and
therefore the resultant radiation spectrum, will display only one break
but its evolution with time will be determined by the dominant cooling process
responsible for the break.

%--------------------------------------------------------------------------------------------%
%--------------------------------------------------------------------------------------------%

Let $\gamma_{sc}$ and $\gamma_{ic}$
be the characteristic Lorentz factors where the distribution of emitting particles 
steepens due to the synchrotron and IC cooling, respectively:
\begin{eqnarray}
\gamma_{sc}	&	=	8.9 \times 10^{3} ~B^{-2} ~t^{-1}_{d} \\	
\gamma_{ic}	&	=	134 ~r^{2}_{15} ~L^{-1}_{42} ~t^{-1}_{d}
\label{eqn:gamma}
\end{eqnarray}
The characteristic synchrotron frequency of emission due to an electron 
of Lorentz factor $\gamma_{e}$ is $\nu_{e}=1.2 \times 10^{6} B \gamma^{2}_{e}$ Hz.
Using this, the ratio of synchrotron and IC cooling break frequencies would be
$(\nu_{sc}/\nu_{ic}) \propto L^{2}_{42}\,B^{-4}\,r^{-4}_{15}$.
We assume that the magnetic field energy density is a constant franction 
($\epsilon_{B}$) of the thermal energy density behind the shock and 
$\epsilon_{B} \approx 0.1$. The thermal energy density behind the shock wave
is given by $U = (9/8) n m_{p} \beta^{2} c^{2}$.
Expressing the CSM density $n$ in terms of mass-loss rate ($\dot{M}$ in units of 
$10^{-6}\,\rm M_{\odot}/yr$ and wind velocity $v_{w}$ in units of $1000\, \rm km/s$) and 
$r =\beta\,ct$ for a freely expanding shock wave, gives the ratio of the break frequencies to be;

\begin{equation}
\left( \frac{\nu_{sc}}{\nu_{ic}} \right) =	2.6 \times 10^{3}\,\frac{L^{2}_{42}}{\beta^{4}_{-1}\,\epsilon^{2}_{B,-1}} \,\left( \frac{\dot{M}_{-6}}{v_{w,3}} \right)^{-2}
\label{eqn:nuc_nuIC}
\end{equation}
It is important to note that this ratio is independent of the source size. 
We can therefore state a general result that \emph{for low to intermediate 
mass-loss rates ($\dot{M} \leq 10^{-5}\, \rm M_{\odot}/yr$)
the dominant cooling process in the radio SNe would be inverse Compton
rather than synchrotron cooling.} 

The only time dependence in equation \ref{eqn:nuc_nuIC} is through the evolution of
photospheric emission expressed as $L_{42}$. The optical peak brightness is typically 
reached around 20 days after the SN. Since $\nu_{sc}\propto t_{d}$, which
follows from Equation \ref{eqn:gamma} and subsequent discussion, it can be shown 
that the synchrotron cooling break is already out of the radio frequency bands
by the time SN is at its peak bolometric luminosity. 
Therefore, the synchrotron cooling break would not be visible in typical radio SNe.

Having demonstrated that the early steep radio SED is likely due to the IC scattering
of optical photons, we now proceed to discuss model fitting and estimation 
of physical parameters for SN 2013df.

\section{Model Fits, Spectral and Physical Parameters}
\label{sec:modeling}
%------------------------------------------------------------------------------------------------------------%
\begin{figure*}
\begin{center}
\includegraphics[width=18cm,clip=False]{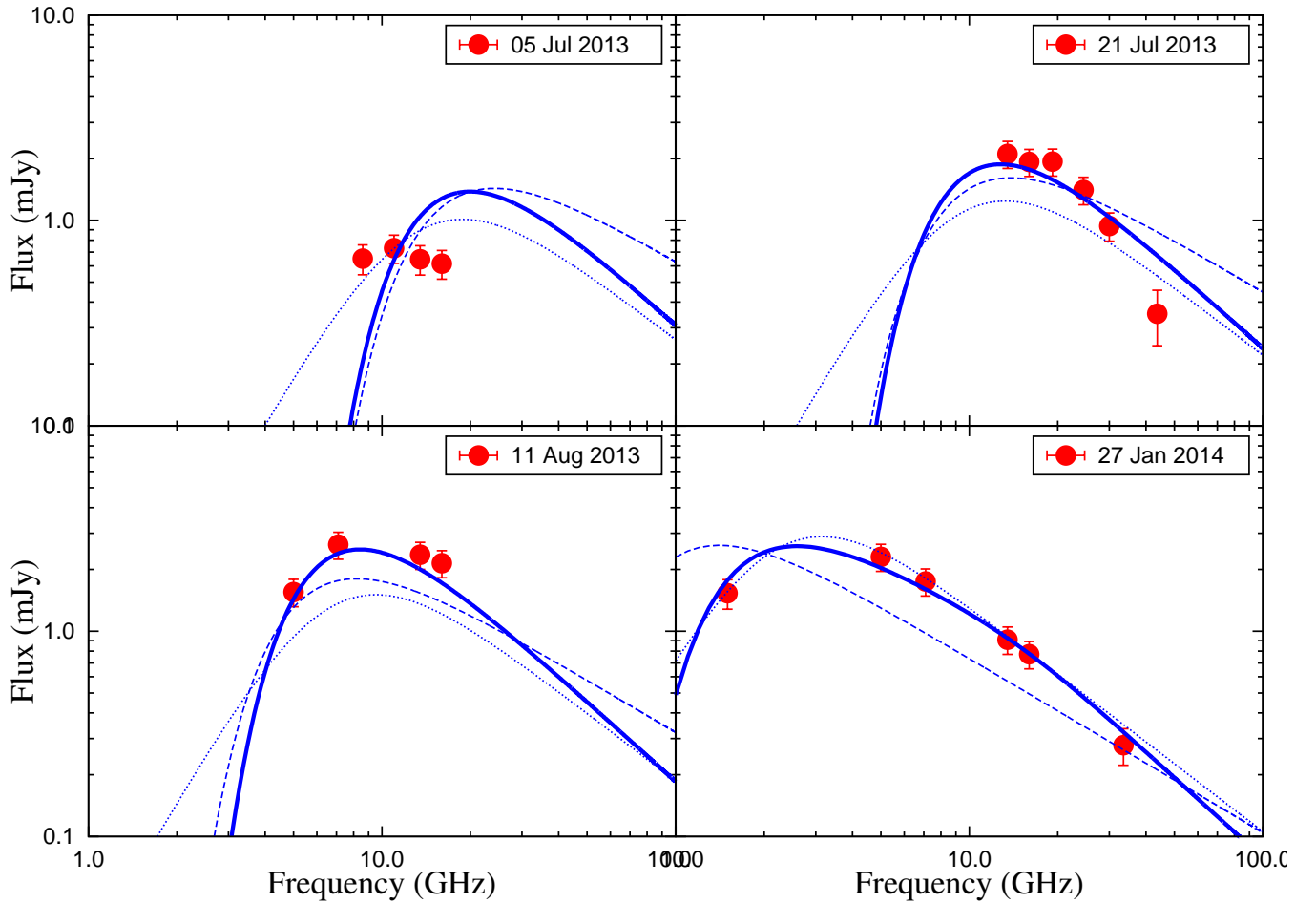}
\caption{Radio SED of SN 2013df at different epochs. The lines indicate the best fit models: 
SSA as the dotted lines, FFA $\otimes$ SSA as the dashed lines and 
IC $\otimes$ FFA $\otimes$ SSA as the solid lines.}
\label{fig:sed}
\end{center}
\end{figure*}
%------------------------------------------------------------------------------------------------------------%

%------------------------------------------------------------------------------------------------------------%
\begin{figure*}
\begin{center}
\includegraphics[scale=0.5,width=18cm]{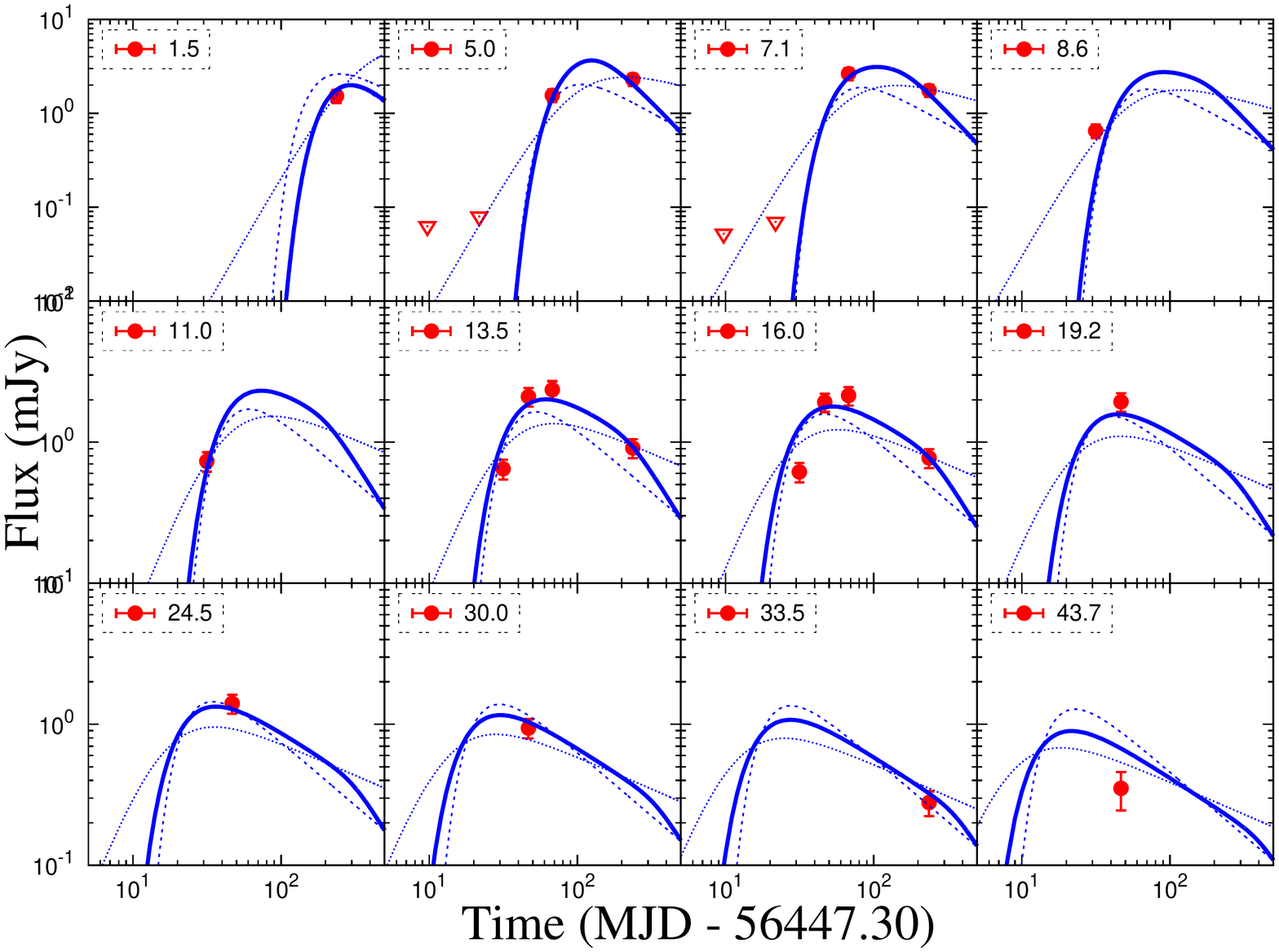}
\caption{Radio light curves of SN 2013df at various radio-frequency bands. The lines indicate the best fit models: 
SSA as the dotted lines, FFA $\otimes$ SSA as the dashed lines and 
IC $\otimes$ FFA $\otimes$ SSA as the solid lines.}
\label{fig:lc}
\end{center}
\end{figure*}
%------------------------------------------------------------------------------------------------------------
	% Model description, fit parameters etc.
	As discussed in the previous section IC cooling modifies the electron distribution
	and consequently the synchrotron emission. Evolution of this electron distribution
	with time depends on the bolometric luminosity as well as the size of the SN.
	In order to parametrize this evolution we have characterized the SED using
	the synchrotron self-absorption break frequency ($\nu_{a}$), 
	the characteristic synchrotron frequency due to electrons with the minimum Lorentz factor ($\nu_{m}$) and the
	inverse Compton cooling break frequency
	($\nu_{ic})$, as well as the spectral peak $f_{\nu_{a}}$ of the synchrotron spectrum.
	In the appendix, we provide spectral and temporal scalings for the evolution 
	of these spectral breaks.
	
	% Fitting procedure : smooth broken power law
	To fit all the radio observations collectively we followed a procedure similar to that 
	of \citet{Kamble2014} and scalings from the Appendix here. By using least square minimization
	we estimated values for the spectral peak $f_{\nu_{a}}$, break frequencies 
	$\nu_{a}, \nu_{ic}$, frequency of free-free absorption $\nu_{\rm ff}$
	where optical depth to free-free absorption $\tau_{\rm ff} = 1$ (for definition 
	see Equation \ref{eqn:tauff} below), 
	electron distribution index $p$ and deceleration parameter $m$. 
	The bolometric luminosity of SN\,2013df was used as an input to determine 
	the evolution of $\nu_{ic}$.
	
	The statistical uncertainties in flux densities measured and listed in Table \ref{tab:vla} 
	are the respective image RMS.
	Variations in the local weather at the telescope can introduce additional errors to 
	the measured image RMSs. Furthermore, calibration of the flux and phase calibrators
	can add further systematic errors of a few percentage. Therefore, during the model fitting, 
	about $15\%$ of the total error was added in quadrature to account for these sources 
	of errors. 
		
	The radio SED indicate that the frequency $\nu_{m}< \nu_{a}$ during all the observations
	and in any case is below the observing band.
	As a result, we can constrain the spectral break $\nu_{m} \ll 1$ GHz.
		
	The best fit value of spectral peak flux density is determined to be 
	$f_{\nu_{a}} = 48 \pm 5$ mJy 
	at the self-absorption frequency $\nu_{a} = 3.3 \pm 0.5$ GHz
	on 21st July 2014 or at $t = 46.8$ days after the burst and the best fit 
	reduced $\chi^{2} = 5.2$ per degrees of freedom (DOF) for 14 DOF. 
	We note that a significant contribution to the $\chi^{2}$ comes from the earliest 
	epoch of radio detections on 05 July 2013. 
	Dropping those observations from the fit results in a significant improvement with reduced 
	$\chi^{2} = 1.7$ for 10 DOF. The resultant model SEDs are compared 
	with observations in Figure \ref{fig:sed} and the light-curves in Figure \ref{fig:lc}.
	
%	The least-square fits, in general, are driven by the data points with smaller error 
%	bars which have larger weights. In addition to that, we emphasize, these are simultaneous 
%	fits to the entire data-set and not separate fits to individual SEDs. This makes the procedure 
%	robust against any individual datapoint possibly straying away from the ensemble.
	
	The Figures \ref{fig:sed} and \ref{fig:lc} highlight the importance of 
	individual emission 
	and suppression processes such as SSA (synchrotron self-absorption), 
	FFA (free-free absorption) and IC (inverse Compton). At early epochs, 
	the model synchrotron emission is under-estimated and can not anticipate the subsequent 
	rise very well. The rising SEDs suggest an accelerating shock wave with $\alpha_{r} = 1.12$ 
	(see Equation\ref{eqn:alphar}). The best fit parameters obtained using the SSA model alone 
	(reduced $\chi^{2}= 3.8$ for 16 DOF), $f_{\nu_{a}} = 1.8 $ mJy and 
	$\nu_{a} = 15.6$ GHz result in an unrealistically slow shock wave $v_{sh}=0.02 \,c$.
	This is significantly slower than the velocities estimated by the emission lines in optical spectra 
	and typical ejecta velocities of $v_{ej} \approx 10,000$ km/s in SNe. 
	This, therefore, clearly indicates that the free-free absorption must be playing 
	an important role in SN 2013df.

	Inclusion of free-free absorption improved the estimates of the shock velocity consistent with the typical values, but not necessarily the quality of the fits (reduced $\chi^{2}= 6.8$ for 15 DOF). The SSA and SSA$\otimes$FFA
	models however result in shallower SEDs which do not agree well with observations
	at early epochs, most notably on 21st July 2013. The steep spectral shape should 
	be related to the underlying steep electron energy distribution and therefore 
	could be the result of a cooling process. As discussed in section $\S$ \ref{sec:IC}
	inverse Compton cooling is likely to play the significant role. 	
	We therefore believe that the steep SED is the result of inverse Compton cooling process.
			
	At late epochs, the radio emission
	at GHz frequencies is expected to be less affected by the suppression effects: 
	a) the synchrotron self absorption ($\nu_{a}$) falls below GHz bands, 
	b) effects of free-free absorption diminishes rapidly with increasing SN size 
	as the optical depth to free-free absorption $\rm \tau_{ff} \propto r^{-3}$ and c) 
	inverse Compton cooling 
	becomes insignificant as the reservoir of optical photons is drained off with 
	diminishing bolometric luminosity. As a result, all three model estimates 
	roughly agree with each other on the last epoch of observations on 27th January 2014
	i.e. about 237 days after the SN.
	
		% Best fit value of p, comparison with other SNe
		For the electron energy distribution index, the fit converged to $p=2.7 \pm 0.3$. The observations before 50 days show significant 	effects of inverse Compton cooling through the steep spectral index $\nu^{-\beta}$	where $\beta = 1.6 \pm 0.2$. As discussed in the appendix this spectral regime 	is related to the electron distribution index through $\beta = p/2$ and is consistent 	with the measured $p=2.7\pm0.3$. At later epochs, when the inverse Compton subsides, optically thin synchrotron spectrum is related to the electron distribution	as $\beta=(p-1)/2$. The observed $\beta=0.9\pm0.1$ is therefore consistent with it.

		The shock wave expansion is best fit as $r=r_{0} (t/t_{0})^{\alpha_{r}}$ with $\alpha_{r} = 0.84\pm0.1$ (see Equation \ref{eqn:alphar}). This is consistent with the expectations for a compact, radiative envelope star where the outer density profile should tend to a power law in radius, $\rho_{sn} \propto r^{-10.2}$ giving $\alpha_{r}=0.88$ \citep{Chevalier1982,Matzner1999}. Our estimate $\alpha_{r}=0.84 \pm 0.1$ is also consistent with a convective outer envelope ($\rho_{sn} \propto r^{-12}$) where one expects $\alpha_{r} = 0.9$.
		 		 
%-------------------------------------------------------------------------------------------%		 
\subsection{X-rays as Thermal Bremsstrahlung emission}
X-ray emission in SNe could be produced by one of the three radiation processes: synchrotron, inverse Compton and bremsstrahlung. Using the radio spectral index of $\beta=-1.6$ based on observations and modeling of the radio emission, we expect the synchrotron emission in X-ray bands to be more than two orders of magnitude fainter as shown in Figure \ref{fig:XraySED}.

Inverse Compton scattering of the photospheric optical emission in to the X-ray bands by relativistic non-thermal electrons behind the forward shock is another possibility. The X-ray spectrum produced by such scattering would be soft. The spectral photon index of the observed X-ray emission in SN 2013df ($\Gamma = 1.3$) is, however, significantly hard and is incompatible with a spectrum dominated by Comptonization. Instead, such a hard spectrum is consistent with the expectations from thermal bremsstrahlung emission.

SN 1993J was another type IIb supernova with bright X-ray emission detected for up to several tens of days. \citet{Fransson1996} modeled it using the thermal bremsstrahlung emission from the forward and reverse shocks that propagate into the stellar wind and ejecta, respectively. Following a detailed analytic argument \citet{Fransson1996} showed that the reverse shock was radiative and  most of the X-ray emission during the first few months originated in the forward shock. They estimated the pre-SN mass-loss rate to be $\sim 4 \times 10^{-5}~\rm M_{\odot}/yr$ for a wind velocity of 10 km/s.

\begin{figure}
\begin{center}
\includegraphics[scale=0.5,width=9cm]{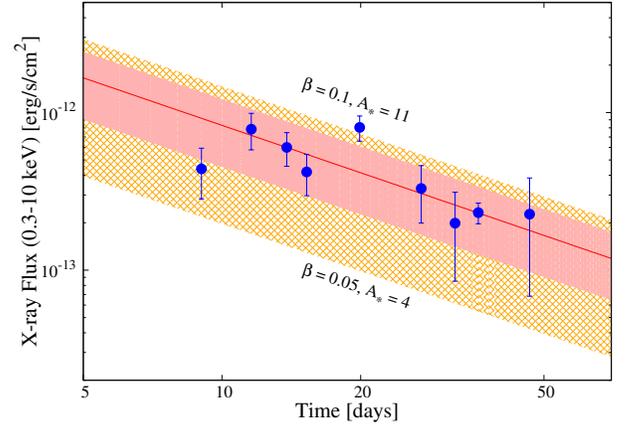}
\caption{Thermal bremsstrahlung emission in X-rays. The \emph{Chandra} and \emph{Swift} observation are shown as blue data points. The modelled thermal bremsstrahlung emission is shown by the shaded region. The lower bound corresponds to the shock velocity $\beta = 0.05 \,c$, mass-loss rate parameter $A_{*} = 4$ and the upper bound to $\beta=0.1, A_{*}=11$.}
\label{fig:LumX}
\end{center}
\end{figure}

Following the procedure outlined in \citet{Fransson1996}, we come to the similar conclusions for SN 2013df as those for SN 1993J. In Figure \ref{fig:LumX} we show a range of mass-loss rates and forward shock velocities to compare the resultant X-ray flux due to thermal bremsstrahlung emission with observations.

As pointed out by \citet{Fransson1996} the X-ray emission may originate either in the forward shock that is sweeping the dense pre-SN wind or in the reverse shock that forms at the contact discontinuity and propagates into the denser ejecta. For homologous expansion, as is approximately the case for SNe, it is convenient to express the ejecta density profile is as a function of velocity, v, as $\rho_{sn}\propto v^{-n}$. The reverse shock temperature may depend on this density profile. Following \citet{Fransson1996}, we find that the temperature of the reverse shock $T_{rs}\geq 10^{7}$ K and that of the forward shock  $T_{fs}\geq 10^{9}$ K for reasonable shock velocities $v_{sh} \geq 10^{4}$ km/s and density profile of the ejecta $n \leq 15$.

If the reverse shock is radiative then a cool shell forms between between the reverse and forward shocked material which absorbs all the X-rays from the reverse shock. For $T > 2 \times 10^{7}$ K, cooling is dominated by bremsstrahlung emission and below that it is dominated by line emission. At such temperatures reverse shock remains radiative for a significant amount of time. For $\dot{M} \geq 8\, \rm M_{\odot}/yr$, $n \geq 12$ and $v_{sh} \leq 21,000$ km/s we find that the reverse shock remains radiative for at least 50 days after the SN which spans the entire duration of our X-ray observations. Thus, all X-rays from the reverse shock are absorbed and most of the observed X-ray flux is due to the forward shocked material.

\emph{Chandra} observation provides the best constraints on the model parameters. The red shaded area in Figure \ref{fig:LumX} corresponds to $3\sigma$ contour bounded by $(6.1 \leq \dot{M} \leq 10.1) \times 10^{-5} \,M_{\odot}$/yr ($v_{w} = 10$ km/s) with the forward shock velocity being ($0.05 \leq \beta \leq 0.1$). The corresponding mass-loss rates for the $5\sigma$ contours shown in larger shaded area are ($4 \leq \dot{M} \leq 11.1) \times 10^{-5} \, M_{\odot}$/yr, $v_{w} = 10$ km/s for similar shock velocities. The solid red line corresponding to the best model estimate for reproducing \emph{Chandra} observations gives $\dot{M} = 8.3 \times 10^{-5} \, M_{\odot}$/yr, $v_{w} = 10$ km/s and $\beta = 0.07$. 

%------------------------------------------------------------------------------------------------------------%
%------------------------------------------------------------------------------------------------------------

\subsection{Model Parameters from Radio Emission}
Another effect of the stellar mass loss is in the form of absorption of radio emission by the ionized wind surrounding the SN. The free-free absorption optical depth of this unshocked gas at frequency $\nu$ is given by 
\begin{equation}
\tau_{\rm ff} \approx 1.9 ~A^{2}_{*} \left( \frac{\rm T}{5\times 10^{4}\, \rm K} \right)^{-3/2} 
  				\left( \frac{\nu}{\rm GHz} \right)^{-2}
  				\left( \frac{\beta}{0.1} \right)^{-3}
  				\left(\frac{\rm t}{25 \,\rm day}\right)^{-3}
\label{eqn:tauff}
\end{equation}
where $\beta$ is the shock velocity in units of the speed of light and and A* is the mass
loss rate in terms of a reference value of $5 \times 10^{13}$ gm/cm. The mass loss parameter $A=\dot{M}/4\pi V_{w} \rm gm~ cm^{-1}$ and our reference value of $5 \times 10^{13}$ gm/cm is attained for $\dot{M} = 10^{-5} M_{\odot}/\rm yr$ and $v_{w} = 10$ km/s.

At $\rm t=46.8 \,day$ after the SN, our best fit model gives $\tau_{\rm ff} = 1$ at $\nu = 10 \pm 1$ GHz. Equation \ref{eqn:tauff} can then be used to get the mass loss rate parameter in terms of shock velocity $\beta$ as $A_{*} = 584\, \beta^{3/2}$.

Following \citet{Kamble2014}, we express the synchrotron frequency and flux in terms of unknown model parameters:
\begin{eqnarray}
\nu_{a}	&	=	&	4.0	\times 10^{11} A_{*}^{0.65} \left( \frac{\beta}{0.1} \right)^{0.72}
   					\left( \frac{\epsilon_{B}}{0.1} \right)^{0.35} 
   					\left(\frac{\epsilon_{e}}{0.1}\right)^{0.51} 
   					\left(\frac{\rm t}{\rm day}\right)^{-1} \rm GHz\\
f_{\nu_{a}}	&	=	&	0.35 A_{*}^{1.37} \left(\frac{\beta}{0.1}\right)^{3.79} 
					\left(\frac{\epsilon_{B}}{0.1}\right)^{0.63} 
					\left(\frac{\epsilon_{e}}{0.1}\right)^{1.27}
					\left(\frac{d_{L}}{100\rm\,Mpc}\right)^{-2}   \rm mJy
\label{eqn:SyncPara}
\end{eqnarray}
where $\epsilon_{e}$ and $\epsilon_{B}$ are the fractions of thermal energy behind the shock that goes into accelerating electrons and magnetic fields, respectively.

Using the best fit spectral parameters determined from radio emission modeling described in the previous section one can simultaneously solve Equations \ref{eqn:SyncPara} and \ref{eqn:tauff} for physical parameters $\beta, \epsilon_{e}$ and $\epsilon_{B}$. Using this approach we estimate the shock velocity to be $\beta = 0.07 \pm 0.01$ which is faster that the velocity of the fastest ejecta estimated from the optical line-widths of about $10,000$ km/s \citep{Morales-Garoffolo2014,Ben-Ami2014,Van Dyk2014, Maeda2015}. This estimate is also consistent with the thermal bremsstrahlung emission observed in X-rays (see Figure \ref{fig:XRTtotal}). For SN\,1993J the velocity estimated from optical lines was about $19,000$ km/s, and a shock velocity was estimated to be $21,600-24,700$ km/s \citep{Fransson1998} in agreement with the VLBI measurements of \citet{Bartel1994,Bartel2002}, which gives $\beta = 0.073$, same as that estimated above for SN\,2013df. The SN mass-loss rate is then estimated to be $\dot{M} = 10.5\pm3.0\, \rm M_{\odot}/yr$ for the wind velocity of $v_{w}=10\, \rm km/s$, in agreement with the estimates from the X-rays. It is reassuring that in an independent approach \citet{Maeda2015} reach roughly similar conclusions using late time optical evolution of SN\,2013df.

Similarly, we determine $\epsilon_{e} = 0.2 \pm 0.02$ and $\epsilon_{B} = (1.0 \pm 0.1) \times 10^{-3}$. While this determined value of $\epsilon_{e}$ is similar to those generally found in SNe, we find a significant deviation from energy equipartition between magnetic fields and accelerated particles. For comparison, \citet{Fransson1998} deduced $\epsilon_{B} = 0.14$ for SN 1993J. The high mass-loss rate combined with low $\epsilon_{B}$ may be the reasons for unusual radio properties of SN 2013df. The only other SN for which inverse Compton effect was shown to be of importance in shaping radio and X-ray emission was SN 2002ap, a type Ic supernova. It is interesting to note that a low value of $\epsilon_{B} = 2 \times 10^{-3}$, similar to the present case of SN 2013df, was deduced for SN\,2002ap also \citep{Bjornsson2004}.

\subsection{Inverse Compton Cooling Break in Radio}
The break in the energy distribution of relativistic electrons due to the IC cooling corresponds to a spectral break $\nu_{IC}$ discussed in section \ref{sec:IC}. Following that discussion, it can be shown in terms of the physical parameters
\begin{equation}
\nu_{IC} = 1.4 \times 10^{10} \frac{\beta^{4}_{-1}}{L^{2}_{42}} \left( \frac{\dot{M}_{-5}}{v_{w,1}} \right)^{-2} t_{d} \sqrt{\epsilon_{B,-1}} \, {\rm Hz}
\end{equation}

Due to its strong dependence on shock velocity, $\nu_{IC}$ favors a slow shock wave and lower mass-loss rates. We obtained $\nu_{IC} = 0.6 \, \rm GHz$ at $t_{0} = 46.8$ days for the best fit to the observations. The bolometric luminosity of SN 2013df at that epoch is $\approx (0.8 \pm 0.3) \times 10^{42}$ erg/s \citep{Morales-Garoffolo2014}. For these observables one can estimate a combination $\beta \epsilon^{2/19}_{B} = 0.01$. For the fastest line velocities $\beta = 0.04$ which gives $\epsilon_{B} = 10^{-5}$. However, due to the strong dependence of $\nu_{IC}$ on other quantities, especially $L_{bol}$ and $\beta$, this comes with a large uncertainty $\epsilon_{B} = (1 \pm 2) \times 10^{-5}$. 

\subsection{Inverse Compton emission in X-rays}
% Introduction
The relativistic electrons which generate the synchrotron radiation and powers the radio emission
in SN shock wave, will naturally up-scatter the optical photons that emerge from the photosphere 
of the SN. If the seed photons or the relativistic electrons have a wide energy distribution, 
the spectrum of the up scattered inverse Compton emission will also be wide. One can, however, 
estimate the total contribution of IC component in a quasi-monochromatic X-ray band. 
The resultant X-ray emission will be related to the energy distribution of electrons. 
The non-thermal electron distribution will produce non-thermal X-rays. In the process of 
IC scattering, the electrons lose their energy. Higher the electron Lorentz factor, faster 
it loses energy to inverse Compton. This, in turn, changes the shape of the electron 
distribution at higher Lorentz factor, making it steeper than the electron distribution injected 
at the shock front.

% Formulation
Radio observations of SNe are an excellent probe of the underlying electron distribution and magnetic field behind the shock. One can construct bolometric lightcurve of SNe using optical observations. Thus, between radio and optical, one has all the necessary information required to understand and estimate the inverse Compton emission. Below we use these observations to self-consistently model the IC emission.

From the previous section, radio observations on 21st July 2013, or at supernova age of $t_{0} = 46.8$ days, indicate $B \approx 0.8$ G. Therefore, the magnetic field energy density behind the shock is $U_{B} = 0.02\, \rm erg/cm^{3}$. Radio observations also indicate the size of the shock wave to be $r=10^{16}\, \rm cm$. From the optical observations we estimate the bolometric luminosity of the SN to be $L_{bol} \approx 0.8 \times 10^{42}$ erg/s on the the same epoch. Thus, energy density in seed photons turns out to be $U_{ph} \approx 0.03\, \rm erg/cm^{3}$. Furthermore, $\xi(p=2.7) = 3.06$ (see Equation \ref{eqn:xi_p} in the appendix), and $f_{\nu_{a}} = 48$ mJy at the self-absorption frequency $\nu_{a} = 3.3$ GHz. Using these values in Equation \ref{eqn:xi_p} and $p=2.7$ as determined from radio observations, we determine $F_{\nu}(1\,\rm keV) = 2.3 \times 10^{-6}$ mJy. This is more than an order of magnitude fainter than the observed X-ray emission. 

Similarly, using Equation \ref{eqn:xi_p} and $p=2.7$, one can estimate the time evolution of IC X-ray flux to be $F_{\nu}(1\,\rm keV) \propto L_{bol}(t)^{0.963} \times t^{-0.925}$. The expected light curve of the IC X-ray emission in 1-10 keV band is significantly fainter. Overall, the contribution of the inverse Compton emission to observed X-ray emission is negligible. 

In other words, the observed X-ray emission is too bright at all epochs than what is expected from the inverse Compton emission. One can therefore conclude that the X-rays from SN\,2013df must be due to another strong emission component. X-ray brightness estimates due to thermal bremsstrahlung, as shown in the previous sections, appear to be reasonable.

\section{Discussion}
\label{sec:discus}

%------------------------------------------------------------------------------------------------------------%
\begin{figure*}
\begin{center}
\includegraphics[width=18cm]{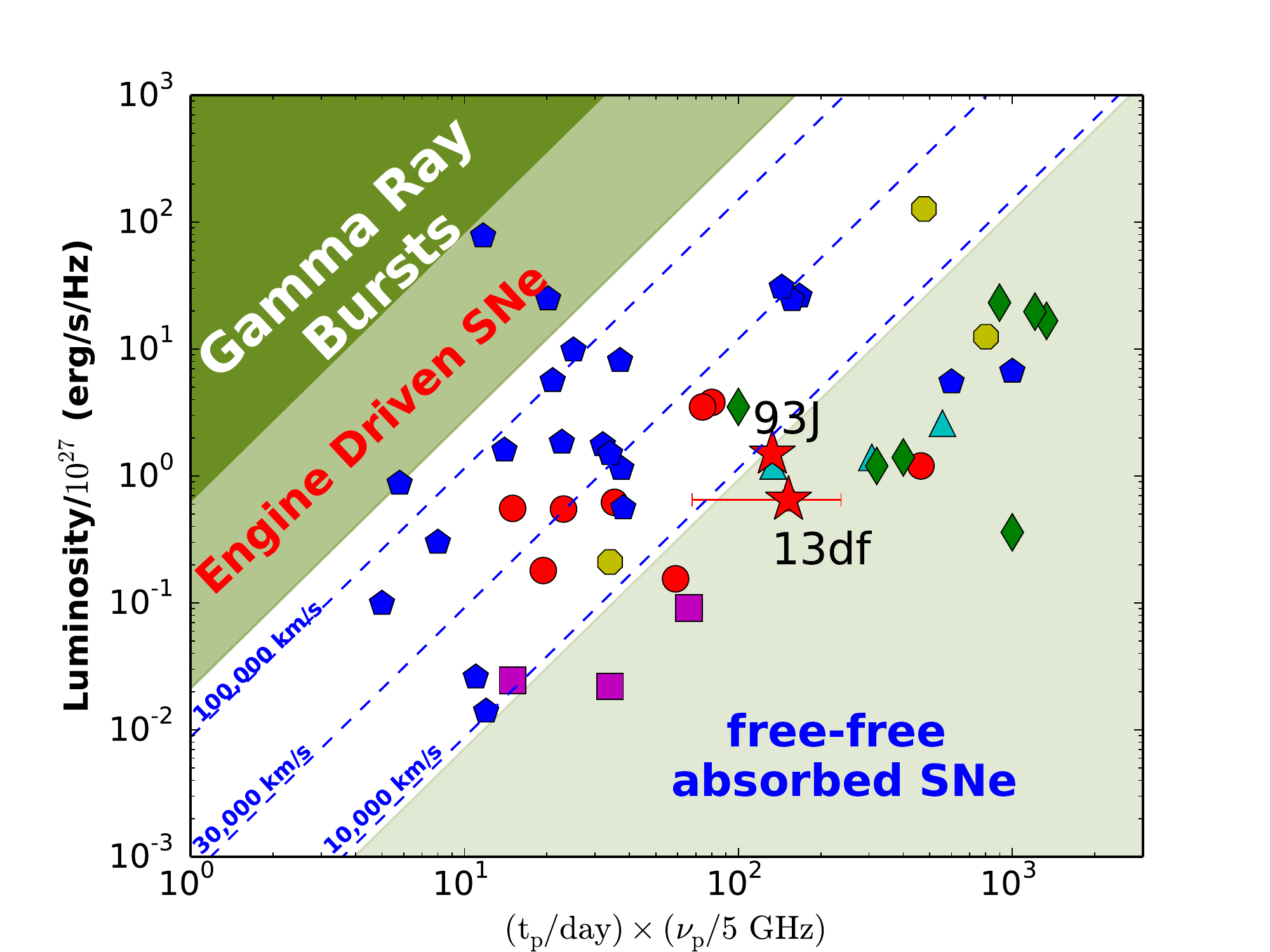}
\vspace{0cm}
\caption{Peak spectral radio luminosity v/s product of the peak time and frequency
for radio SNe of various types - Ib/c: blue pentagons; IIb: red circles; 
IIP: magenta squares; IIn: green diamonds; IIL: cyan triangles and 
II (SN 1978K, 1981K \& 1982aa): yellow octagons. 
The similarity between SN\,1993J and SN\,2013df is clearly evident in this plot.
}
\label{fig:CPlot}
\end{center}
\end{figure*}
%------------------------------------------------------------------------------------------------------------%
\subsection{Radio SNe - a global view}

% Intro
It is instructive to view the SN 2013df in comparison with radio SNe in general. 
The peak Luminosity v/s peak time plot offers the most informative way to project 
the radio SNe. The positions of radio SNe in such a plot are shown in Figure \ref{fig:CPlot},
which is an update of the previous plots in \citet{Chevalier1998,Chevalier2006}.
Different SN types have been color coded for the ease of distinction 
- Ib/c: blue pentagons; IIb: red circles; 
IIP: magenta squares; IIn: green diamonds; IIL: cyan triangles and 
II (SN 1978K, 1981K \& 1982aa): yellow octagons.

% Assumptions
While both the co-ordinates of SNe on this plot are purely observables, 
the underlying lines depend, though weakly, on certain model 
assumptions. The blue dotted lines of constant shock velocity assume energy 
equipartition between synchrotron emitting electrons and magnetic field, 
and that there is no strong cooling of the electrons. The lines also assume
a uniform electron energy distribution index of $p=3$. Despite these assumptions
the plot provides an easy way of comparison between different SNe as well as
insights into different classes of SNe and their progenitors.

The slowest 
to fastest moving objects appear from right to the left of the plot. 
The blue dotted lines indicate mean velocities of the radio shocks if SSA 
is the responsible for the spectral peak. The relativistic synchrotron sources
such as gamma ray bursts appear to the far left of the plot in the dark green shaded area.
The relativistic SNe, such as SN 2009bb, appear in the light green shaded area 
labelled as the engine driven SNe. As one moves to the right 
of the plot synchrotron emission is increasingly suppressed by the FFA 
in the progenitor wind.

% FFA line
The effect of FFA is often noticeable in radio observations where the SED falls 
faster than in the SSA towards low radio frequencies. Consequently,
the rise of radio light curve is also faster at early times in the presence of FFA.
Our early non-detections of SN 2013df at C band clearly bring out the presence 
of FFA. Ignoring FFA results in the estimates of reduced shock-velocity, 
which often fall short of photospheric velocities measured from optical 
line emission. Therefore, we use nominal a velocity of $< 10,000$ km/s 
as a distinction between FFA and SSA region in the luminosity plot 
(Figure\,\ref{fig:CPlot}).

\subsection{Progenitors of SNe IIb - extended versus compact}
\label{sec:progenitors}

Figure \ref{fig:CPlot} captures the essence of radio emission in SNe and portrays 
the diversity among various types of SNe. Radio SNe show a wide luminosity 
distribution spread over more than four orders of magnitude. Luminosity distribution 
of IIb SNe is significantly narrow, confined within two orders of magnitude. 
But their distribution of peak times is relatively as wide as those of SNe Ibc. 
The distribution merges with those of SN Ib/c.
As one moves to the right, the effect of FFA becomes increasingly noticeable 
in SNe IIb as has been clearly seen in SN\,1993J \citep{Fransson1998}, 
2013df (this work), 2011hs \citep{Bufano2014} and 2010P 
\citep{Romero-Canizales2014}. 

It is, therefore, plausible that the broad distribution in rise times 
of SN IIb may be an outcome of different intrinsic properties of 
the progenitors or the circumstances surrounding the SN.
Indeed, such a proposal has been put forward by \citet{Chevalier2010}
based on distinct emission properties observed in radio, optical and X-rays.

% early rapid optical emission
One clue to this might come from early rapidly evolving optical emission
often termed as the `cooling envelope emission'. Current understanding
of this emission is that following the shock break-out from stellar 
surface, outer envelope of the star expands and cools. 
As the expanding photosphere reaches this stellar envelope the
radiation stored in the envelope manages to escape. This `hot' radiation 
results in the rapidly evolving UV/optical emission at early times lasting
from a few hours to several days. The cases where such emission 
has been detected are only a few consisting of SNe of type IIb: 1993J, 2011dh, 
2011hs, 2011fu and most recently 2013df, and type Ic: SN 2006aj 
and a few cases of SN of type IIP.

This however is not a complete picture. Observations are maturing 
rapidly in discovering SNe early, in carrying out rapid follow up 
and in UV capabilities.
Rapid follow up reveals that SNe does not always develop such double 
peaked early optical emission. Despite observations at very early times, 
several SNe have been detected which do not show early optical 
emission to very deep limits which might be problematic 
to the theories of cooling envelope emission, especially if the SNe 
are expected to have similar progenitors.
It may, therefore, be insightful to tie up such properties with multiband 
long term follow up observations.

%------------------------------------------------------------------------------------------------------------%
\begin{figure*}
\begin{center}
\includegraphics[scale=0.8,width=8cm,height=6cm]{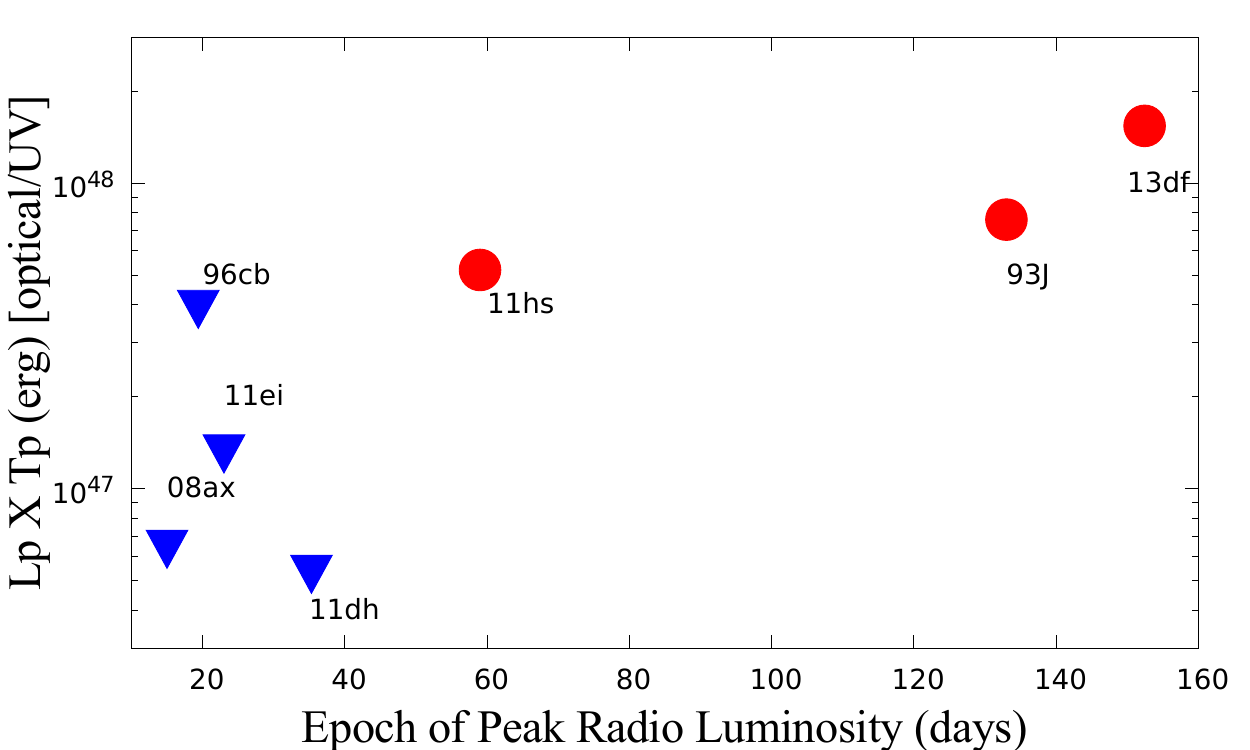}
\includegraphics[scale=0.44]{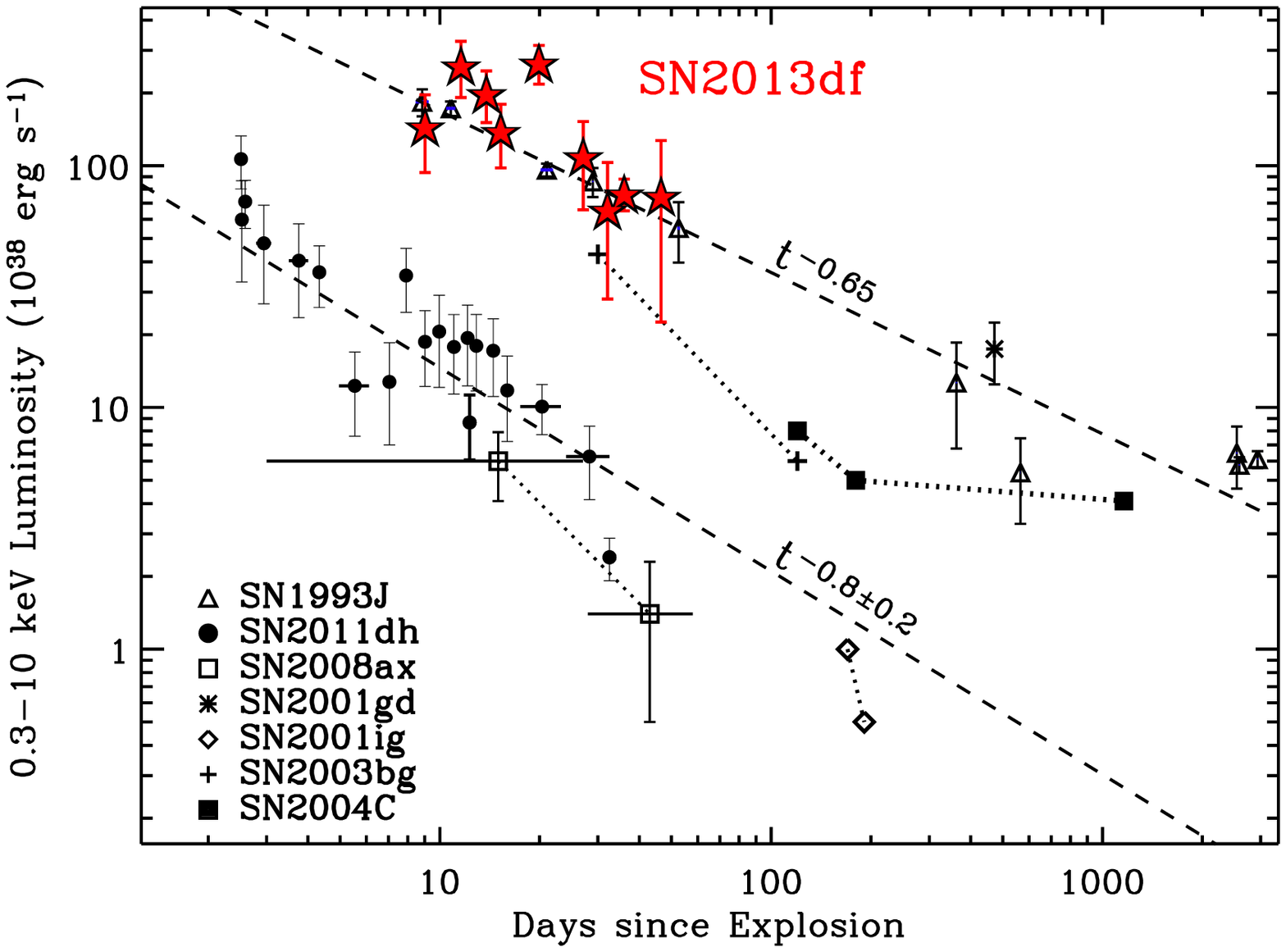}
\vspace{0cm}
\caption{{\bf Early optical emission and radio properties:} A product of the early peak 
bolometric luminosity and its epoch are plotted against the epoch of radio peak 
brightness (Left panel). The quantity on the y-axis is a measure of the amount of energy
a SN deposits in the envelope. The red points are the SNe with detected 
early optical emission. The blue upper limits are the SNe where early 
UV/optical observations have been reported without similar long lasting 
optical emission. The detections and non-detections of early optical peak
are clearly separated signaling apparent dichotomy.
{\bf X-ray emission in type IIb:} Compared to a sample of type IIb SNe, SN\,2013df 
shows luminous X-ray emission with strong similarities to SN\,1993J, both in terms
of luminosity and temporal decay rate (Right panel). The SNe with bright X-ray emission 
are also the ones which show FFA and delayed peak brightness in radio.
References: SN\,1993J \citep{Chandra2009}, SN\,2001gd  \citep{Perez-Torres2005}, 
SN\,2003bg \citep{Soderberg2006}, SN\,2004C (Dittman in prep.), 
SN\,2008ax \citep{Roming2009}, SN\,2011dh  \citep{Soderberg2012}.
}
\label{fig:RadioOptical}
\end{center}
\end{figure*}
%------------------------------------------------------------------------------------------------------------%

A simple but surprising trend seem to emerge between 
this early optical emission and slowly evolving SNe. And at the moment
we concern ourselves with of those of type IIb only. 
All SNe in the FFA region of Figure \ref{fig:CPlot}
exhibit this early optical emission whereas others do not ! 
As mentioned before, SNe 1993J, 2011hs, 2013df are the prime 
examples of this optical emission, they all show strong FFA 
absorption in the radio emission and reach radio peak brightness
comparatively late ($\geq$ 100 days). SNe 2010P which also
falls in the same FFA category was never observed in UV/optical
so we can not comment on that. Opposite of this is SN 2011fu, 
which exhibited the early optical peak but was never followed up in radio
so it is not possible to comment if it also had a strong FFA component.
On the other hand, SNe without strong FFA do not show this 
optical emission to deep limits despite early observations 
(SNe 1996cb, 2008ax and 2011ei). Early UV/optical observations 
have not been reported for SN 2001gd and 2001ig.
SN 2011dh is an interesting intermediate case, where the
early optical emission was detected in one band and was significantly 
short lived. Also, its contribution to the bolometric luminosity is negligible.
The progenitor of this SN has however been identified 
and we discuss that issue later.

These findings are put in persecutive and more quantitatively in 
Figure \ref{fig:RadioOptical} where we plot the product of the early 
peak bolometric luminosity and its epoch against the epoch of radio 
peak brightness.
The right panel of Figure \ref{fig:RadioOptical} shows X-ray emission in SNe IIb.
Interestingly, the bright X-ray emission correlates with radio properties 
such as FFA, delayed radio peak, and double peaked optical emission.
Rapidly evolving radio SNe are also the ones which display faint or none X-ray 
emission. Unlike SN\,1993J and 2013df, X-ray emission of SN 2011dh was 
non-thermal in nature and more than an order of magnitude fainter.

Early UV/Optical observations of SNe provide interesting diagnostics about 
the SN progenitor, especially in cases where pre-SN images of 
the progenitor do not exist. But before that the physical theories 
need to be cross checked and calibrated against the cases 
where the progenitor has been identified. Undoubtedly, such detections 
will improve in future and we will have more candidates 
to test theories against. At the moment we have four SNe with identified
progenitors, properties of which we discuss in the next section. 

\subsection{SNe IIb with identified progenitors}
Observations are not only getting quick in responding to the SNe fast enough to catch the early emission but also probing deeper to match and identify the progenitors from archival images of nearby SNe. As a result progenitors of four SNe of type IIb have been identified in pre-SN images. A YSG of radius $\approx 600\,R_{\odot}$ was identified as the progenitor of SN 1993J. A similarly large YSG star has been identified as a progenitor of SN\,2013df. A less extended, a YSG nevertheless, with $R\approx 200 R_{\odot}$ was found to have disappeared from the position of SN\,2011dh and has therefore been concluded as the progenitor of SN\,2011dh. With the surge in direct identification of SN progenitors it is now becoming increasingly possible to test some of the  predictions about early emission properties of similar SNe.

% Why similar types of progenitors should have such distinct properties ?
The question then is why should similar kind of progenitors have different 
emission properties such as in the cooling envelope which emerge from 
close to the stellar surface ? As pointed out before
not every type IIb SN shows cooling envelope emission, and there appears
to be correlation with the progenitor size. Indeed, SN 2008ax did not display
the cooling envelope emission. It was difficult to fit either a single or
binary supergiant or WR + supergiant system self-consistently as progenitors 
to the observed colors from the pre-SN images (Li et al. 2008; Crockett et al. 2008).
The possible initial mass range for the progenitor turns out to be large, 
$\approx 10 - 28 \,M_{\odot}$.

More progress is required on theoretical front in figuring out all the elements 
for the CEE theory to be successful. It has been pointed out by \citet{Nakar2014}
that standard stellar density profile does not lead to CE emission. 
Instead, extended progenitors with non-standard density profiles are necessary. 
Such modifications of density profiles in the extended envelopes may occur only 
in a subset of progenitors due to binary interaction, non-standard mass-loss, 
episodic ejections of outer envelopes and subsequent evolution etc. There are, however, no direct
means of observationally confirming if such modified density profiles actually
exist elsewhere and more so under a given situation. Secondly, current theories do not
take into account detailed dynamics or radiation transport as pointed out 
by \citet{Rabinak2011,Nakar2010,Nakar2014}. Different kinds of progenitors, 
such as WR, BSG and RSG, have been considered by \citet{Nakar2010, Nakar2014}
to estimate CE emission properties. One difference that is expected, and to which 
calculations from different groups agree, is that larger progenitor results 
in longer lasting CE emission.

% Is this really the cooling envelope emission ?
% Must explain the apparent contradictions and complexities.
A credible theory must explain all the apparent contradictions and complexities
as pointed out above. This raises an important question: is the double peaked 
optical light curve really due to the CE emission ? Alternative ideas should 
be explored in order to explain observations satisfactorily. 
Interaction of shock-wave with binary companion is expected to leave
some signature on the optical light curves. Such an idea has been explored
for SN Ia where a white-dwarf accretes matter from the companion ultimately 
leading to the SN. As the SN shock-wave interacts with the envelope of 
the companion dissipating kinetic energy and heating it up, 
it results in the emission that is expected to peak in UV bands on timescales 
of a few days which is roughly similar to that observed in some type IIb. 
It is of considerable interest that a binary companion has also been either 
confirmed (SN 1993J, SN 2011dh) or speculated to the progenitors of these SNe.

% IIb are best candidates for studying extended emission, radio properties, 
% and  progenitors collectively
SNe of type IIb thus provide the best opportunities to study variety of phenomena: 
double peaked bolometric light curves and bright radio and X-ray emission. Due to the 
relatively higher rate of type IIb compared to Ibc they are also best suited
for direct detection of progenitors. With the possibility of such rich datasets
they are ideal for tying together various observed properties and in building
a detailed understanding. Correlating progenitors with radio properties may
be more subtle than the simple dichotomy of compact and extended progenitors
and may be made more complex by the processes that affect stellar evolution
such as stellar rotation, magnetic fields, binary interactions and 
evolution in the late stages \citep{Smith2015}. A more complete picture
can be built by multiband observations and direct detection of progenitors 
in the pre-SN images.

\section{Summary and Conclusions}
We presented extensive radio and X-ray observations of SN\,2013df in the context of type IIb SN explosions. We also presented a model of inverse Compton cooling and estimated its effects on the evolution of radio emission from SN\,2013df.
Our main results can be summarized as follows:
\begin{itemize}
\item 
With bolometric luminosities within a factor of a few of each other, and similarly extended luminous cooling envelope emission and comparable mass-loss rate estimates from the radio and X-ray emissions, SN\,2013df appears similar to that of SN\,1993J, in several ways. This close association is portrayed most easily in Figure \ref{fig:CPlot}. 
Indeed, the progenitor of SN\,2013df appears to have been similar to that of SN\,1993J in mass and radius \citep{Van Dyk2014}.
\item SN\,2013df showed a peculiar soft-to-hard temporal evolution of the optically thin part of the radio spectrum. We interpret this behavior as  free-free absorbed synchrotron radiation where the underlying electron distribution responsible for the radio emission is modified by Inverse Compton cooling at early times (ref to Fig. 3). This model provides a reasonably good match to the observations for $t>40$ days and leads to the picture of a shock wave propagating with modest velocity $v_{sh}=0.07c$ into a dense environment, previously shaped by substantial mass loss from the progenitor star $\dot M=8\times 10^{-5}\,\rm{M_{\odot}/yr}$ (for wind velocity $v_w=10\,\rm{km\,s^{-1}}$). Comparable mass-loss rate was inferred for the type IIb SN\,1993J \citep{Fransson1996}. 
\item Self-consistent modeling of the radio emission of SN\,2013df that includes synchrotron emission and self absorption, free-free absorption and inverse Compton cooling effects shows  significant deviation from the generally assumed equipartition of shock energy between magnetic fields and accelerated electrons. For SN\,2013df we infer  $\epsilon_e\sim 200 \epsilon_B$. 
\item In close similarity to SN\,1993J, the X-ray emission from SN\,2013df  (Fig. 7) is in clear excess with respect to the synchrotron model that best fits the radio emission (ref Fig. 2). The hard X-ray spectrum is dominated by  thermal bremsstrahlung emission with $T>5$ keV and indicates a mass-loss rate $(4<\dot M<11)\times 10^{-5}\,\rm{M_{\odot}/yr}$, consistent with our results above.
\item Our modeling of the radio and X-ray radiation from SN\,2013df further demonstrates the importance of the IC cooling process in shaping the electron distribution responsible for the detected radio emission in SNe at early times.
\item Finally, we observationally link the presence of bright cooling envelope emission in the optical/UV bands at very early times in type IIb SNe to the later appearance of heavily free-free absorbed radio emission (Fig. 7), coupled with hard X-ray radiation from thermal bremsstrahlung (and hence large  mass-loss rate of the progenitor star before exploding). 
The dichotomy of such emission properties displayed by type IIb SNe suggest diversity among progenitors.

\item The current small number of type IIb SNe well observed in the optical/UV, radio and X-ray bands both at early and at late times (Fig. 7) does not allow us to quantitatively test our conjectures. Clearly, a significantly larger sample of type IIb SN explosions with extended multi-wavelength coverage is needed. The case of SN\,2013df demonstrate that multi-band observations provide crucial diagnostics about the nature of SN progenitors. SNe of type IIb are advantageous to carry out multi-faceted investigations including detections of progenitors in pre-SN observations, double peaked optical emission and non-thermal emission.
\end{itemize}

\acknowledgments
Support for this work was provided by the David and Lucile Packard Foundation
Fellowship for Science and Engineering awarded to AMS. One of the authors, RAC, 
acknowledges NASA grant  NNX12AF90G.
We thank Enrico Ramirez-Ruiz, Nathan Sanders and Lorenzo Sironi
for helpful discussions.
{}

\clearpage
\appendix

\section{Synchrotron Emission from Inverse Compton Cooled Electrons in Young Supernovae}
\label{sec:ICappendix}
Synchrotron spectrum produced by the relativistic electrons behind
a non-relativistic shockwave expanding into the surround medium
of density profile $n(r) \propto r^{-2}$ could be described as a broken power law with
two characteristic frequencies ($\nu_{m}, \nu_{a}$) and the spectral peak 
$F_{m}$. Following \citet{Soderberg2006} and \citet{Kamble2014} we express 
the size and speed of the shockwave evolving in time as
\begin{eqnarray}
r	&	=	&	r_{0} \times (t/t_{0})^{\alpha_{r}}\\
\beta	&	=	&	\beta_{0} \times (t/t_{0})^{\alpha_{r}-1}
\label{eqn:alphar}
\end{eqnarray}
where the shock speed is $\beta c = dr/dt$.

\subsection{Spectral shape, Break Frequencies and Radio Light curves}
We identify the spectral break due to inverse Compton 
cooling of electrons as a characteristic frequency ($\nu_{ic}$).
The evolution of this break frequency depends not only  
on the SN size but also on its bolometric luminosity, $L_{bol} \propto t^{\alpha_{L}}$.
Below we describe two spectral regimes that are most relevant for the evolution of radio SNe.

\subsubsection{Case 1: $\nu_m \ll \nu_a \ll \nu_{ic}$}
In  Case 1,  the  characteristic synchrotron  frequency  is below  the
self-absorption  frequency,  $\nu_m \ll  \nu_a$ and therefore the synchrotron emission
is suppressed due to the self-absorption. As a result,  the
spectral  peak occurs at  $\nu_{p}\approx  \nu_a$ and the scalings simplify to 

\begin{equation}
\nu_{p}\approx \nu_a\propto t^{-\frac{(5p-2)-2\alpha_{r}(2p-3)} {p+4}},~~~
f_{\nu_{p}}\approx f_{\nu_a}\propto t^{-(1-\alpha_{r})\frac{12p-7}{p+4}},~~~
\nu_{ic}\propto t^{4\alpha_{r}-\alpha_{L}-3},~~~
\label{eqn:peak_1}
\end{equation}

\noindent
The temporal and frequency dependence of flux $f_{\nu}$
at observed frequency $\nu$ is then generalized by

\begin{equation}
f_{\nu} \propto 
  \begin{cases}
    \nu^{5/2}~t^{2\alpha_r + 1/2} & \nu_m < \nu < \nu_a \\
    \nu^{-(p-1)/2}~t^{-[(5p-3)/2 - \alpha_r(2p-1)]} & \nu_a < \nu < \nu_{ic} \\
    \nu^{-p/2}~t^{-[5p/2 - \alpha_r(2p+1) + \alpha_{L}]} & \nu_{ic} < \nu
  \end{cases}
\end{equation}
\label{eqn:case_1}

\subsubsection{Case 2: $\nu_m \ll \nu_{ic} \ll \nu_{a}$}
In  Case 2,  the  synchrotron self-absorption frequency is above 
the characteristic inverse Compton frequency, $\nu_{ic} \ll  \nu_{a}$. 
Since inverse Compton effect dominates around the epoch of peak bolometric brightness 
this situation is likely to occur around the peak optical brightness of the SN. 
The synchrotron spectral peak would still be at $\nu_{a}$ and the scalings are given by 

\begin{equation}
\nu_{p}\approx \nu_a\propto t^{-\frac{5p-1} {p+5} (1-\alpha_{r}) -\alpha_{r}},~~~
f_{\nu_{p}}\approx f_{\nu_a}\propto t^{\frac{14p+5} {p+5} \alpha_{r} -\frac{13p}{p+5} +\alpha_{L}},~~~
\nu_{ic}\propto t^{4\alpha_{r}-\alpha_{L}-3},~~~
\label{eqn:peak_2}
\end{equation}

\noindent
The temporal and frequency dependence of flux $f_{\nu}$
at observed frequency $\nu$ is then generalized by

\begin{equation}
f_{\nu} \propto 
  \begin{cases}
    \nu^{5/2}~t^{4\alpha_{r}-\alpha_{L}-1/2} & \nu_m < \nu < \nu_{ic} \\
    \nu^{5/2}~t^{4\alpha_{r}-\alpha_{L}-1/2}  & \nu_{ic} < \nu < \nu_{a} \\
    \nu^{-p/2}~t^{-[5p/2 - \alpha_r(2p+1) + \alpha_{L}]} & \nu_{a} < \nu
  \end{cases}
\end{equation}
\label{eqn:case_2}

% A small appendix to explain inverse compton effect in SNe

% explain all the break frequencies and their time dependeces

% Flux evolution in time

\section{Inverse Compton Scattering of photospheric emission}

\begin{figure}
\begin{center}
\includegraphics[scale=0.65]{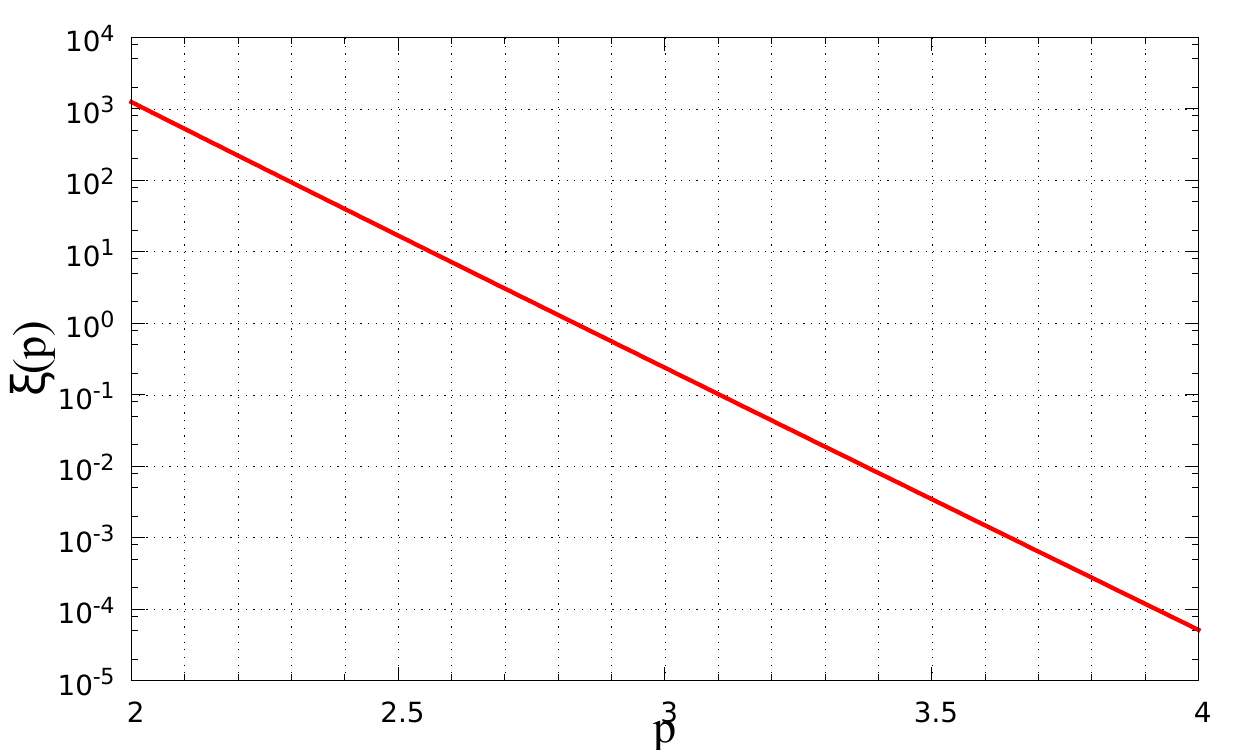}
\caption{The function $\xi(p)$.}
\label{fig:xi_p}
\end{center}
\end{figure}

The same relativistic electrons which emit synchrotron emission are also responsible for the up-scattering of photospheric emission dominated by IR/optical/UV photons. If the scattering electrons have the power law energy distribution ($N(\gamma_{e}) \propto \gamma^{-p}_{e}$), the up-scattered IC radiation carries the similar form ($E^{-(p-1)/2}$). The situation under consideration, IC scattering of Planck distributed photons by relativistic power-law electrons, has been considered in detail by \citealt{Rybicki1979}. Following equation 7.31 and 6.36 of \citet{Rybicki1979} 
we get a simple relation
\begin{equation}
\frac{P_{\nu,syn}}{P_{\nu,X}} = 
	\left( \frac{\nu_{Syn}}{\nu_{X}} \right)^{-\frac{p-1}{2}}
	\frac{U^{\frac{p+1}{4}}_{B}}{U^{\frac{p+5}{8}}_{ph}} \times \xi(p)
\label{eqn:xi_p}
\end{equation}
where we have collected the natural constants $h,c,q,m_{e}$ and dependence on parameter $p$ inside the function $\xi(p)$. We have plotted this function in Figure \ref{fig:xi_p}. The frequency $\nu_{X}$ is the frequency of the up-scattered photons. We use the subscript \emph{X} to distinguish it from the inverse Compton cooling break $\nu_{ic}$ discussed in the previous section.

For $p=3.0$, Equation \ref{eqn:xi_p} reduces to the well known result 
\begin{equation}
\frac{\nu_{syn}\times P_{\nu,syn}}{\nu_{X}\times P_{\nu,X}} \approx \frac{U_{B}}{U_{ph}}
\label{eqn:xi_p3}
\end{equation}
where factor $\xi(p)$ is often neglected. 

The essence of Equation \ref{eqn:xi_p} is that the knowledge of electron distribution and seed photon field, from radio and optical observations respectively, is sufficient to describe the inverse Compton emission completely. For example, it follows from Equation \ref{eqn:xi_p} that for $p=3.0$ the IC flux is
\begin{equation}
F_{\nu,X} = \frac{1}{0.24} \,F_{\nu,syn} \left(\frac{\nu_{X}}{\nu_{syn}} \right)^{-1} \frac{U_{ph}}{U_{B}}
\label{eqn:xi_p3}
\end{equation}
where $U_{B} = B^{2}/8\pi$ and $U_{ph} = L_{bol}(t)/4\pi r^{2} c$. The ratio of energy densities $U_{ph}/U_{B} \propto L(t)$ in the simple case of freely expanding shockwave ($\alpha_{r} = 1$), since $B \propto 1/r$ in the wind medium. Optically thin synchrotron flux in a given band $F_{\nu,syn} \propto t^{-1}$. This gives the scaling of inverse Compton flux as
\begin{equation}
F_{\nu,X} \propto L(t) \times t^{-1}
\label{eqn:Fic_t}
\end{equation}
similar to that given by \citet{Bjornsson2004,Chevalier2006}. From Equation \ref{eqn:xi_p} we can now show that for other values of \emph{p},
\begin{equation}
F_{\nu,X} \propto L(t)^{\frac{p+5}{8}} \times t^{-\frac{p+1}{4}}
\label{eqn:Fic_t}
\end{equation}

%------------------------------------------------------------------------------------------------------------%
%------------------------------------------------------------------------------------------------------------
\end{document}